\documentclass{emulateapj}
\usepackage{graphics,epsfig}

\shorttitle{GTC/OSIRIS Monochromatic Images}
\shortauthors{Mayya et al.}

\newcommand{\ktf}{$\kappa$}
\newcommand{\hii}{\textsc{H\,ii}}
\newcommand{\nii}{\textsc{N\,ii}}
\newcommand{\sii}{\textsc{S\,ii}}

\begin{document}
\centerline{                          .                        }
\centerline{                             To appear in PASP (August 2012)}

\title{
Flux calibrated emission line imaging of extended sources using GTC/OSIRIS 
tunable Filters\footnote{Based on observations made with the Gran Telescopio
Canarias (GTC), installed in the Spanish Observatorio del
Roque de los Muchachos of the Instituto de Astrof\'{i}sica de Canarias,
in the island of La Palma.}
}

\author{   Y. D. Mayya\altaffilmark{1,6},
           D. Rosa Gonz\'alez\altaffilmark{1},
           O. Vega\altaffilmark{1},
           J.~M\'endez-Abreu\altaffilmark{2,3},
           R. Terlevich\altaffilmark{1,4},
           E. Terlevich\altaffilmark{1,5},
           E. Bertone\altaffilmark{1},
           L. H. Rodr\'iguez-Merino\altaffilmark{1},
           C. Mu\~noz-Tu\~n\'on\altaffilmark{2,3},
           J. M. Rodr\'iguez-Espinosa\altaffilmark{2,3},
           J. S\'anchez Almeida\altaffilmark{2,3},
           J. A. L. Aguerri\altaffilmark{2,3}}
\altaffiltext{1}{Instituto Nacional de Astrof\'isica, \'Optica y Electr\'onica, Tonantzintla, Puebla, C.P. 72840, Mexico.}
\altaffiltext{2}{Instituto de Astrof\'\i sica de Canarias, E-38205 La Laguna, Tenerife, Spain.}
\altaffiltext{3}{Departamento de Astrof\'\i sica, Universidad de La Laguna, Tenerife, Spain.}
\altaffiltext{4}{IoA, Madingley Rd., Cambridge CB3 0HA, UK.} 
\altaffiltext{5}{Visiting Fellow, IoA, Madingley Rd., Cambridge CB3 0HA, UK.} 
\altaffiltext{6}{ydm@inaoep.mx}

\begin{abstract}

We investigate the utility of the Tunable Filters (TFs) for obtaining 
flux calibrated emission 
line maps of extended objects such as galactic nebulae and nearby galaxies,
using the Optical System for Imaging and low Resolution Integrated 
Spectroscopy (OSIRIS) at the 10.4-m {\it Gran Telescopio Canarias} (GTC). 
Despite the relatively large field of view of OSIRIS ($8^\prime\times8^\prime$),
the change in wavelength across the field ($\sim80$\,\AA) and the
long-tail of the TF spectral response function are 
hindrances for obtaining accurate flux calibrated emission-line maps of 
extended sources. 
The purpose of this article is to demonstrate that emission-line maps useful
for diagnostics of nebula can be generated over the entire field of view of 
OSIRIS, if we make use of theoretically well-understood characteristics of TFs.
We have successfully generated the
flux-calibrated images of the nearby, large late-type spiral galaxy M101 in 
the emission lines of H$\alpha$, [\nii]$\lambda6583$, 
[\sii]$\lambda6716$ and [\sii]$\lambda6731$. 
We find that the present uncertainty in setting the central 
wavelength of TFs ($\sim$1\AA), is the biggest source of error in the 
emission-line fluxes. By comparing the H$\alpha$ fluxes 
of H{\sc ii} regions in our images
with the fluxes derived from H$\alpha$ images obtained using narrow-band
filters, we estimate an error of $\sim11$\% in our fluxes.
The flux calibration of the images was carried out by fitting
the Sloan Digital Sky Survey (SDSS) $griz$ magnitudes of in-frame stars 
with the stellar spectra from the SDSS spectral database. 
This method resulted in 
an accuracy of 3\% in flux calibration of any narrow-band image, which is
as good as, if not better than, that feasible using the observations of 
spectrophotometric standard stars.
Thus time-consuming calibration images need not be taken. 
A user-friendly script under the IRAF environment was developed and
is available on request.

\end{abstract}


\section{Introduction}
Optical System for Imaging and low Resolution Integrated Spectroscopy (OSIRIS)
is an imager and spectrograph for the optical wavelength range, located 
in the Nasmyth-B focus of the 10.4-m {\it Gran Telescopio Canarias} (GTC)
\citep{Cepa05, Cepa10}.
Apart from the standard broad-band imaging and long-slit spectroscopy,
it provides additional capability such as narrow-band tunable 
filter (TF) imaging, charge-shuffling and multi-object spectroscopy. 
OSIRIS covers the wavelength range from 0.365 to 1.05~$\mu$m with an 
unvignetted field of view (FoV) of $7.8\times7.8$\,arcmin$^{2}$, for direct 
imaging. Narrow-band imaging is made possible through the use of a TF, 
which is in essence a low resolution Fabry-Perot etalon. The filter is tuned 
by setting a specific separation between the optical plates of the Fabry-Perot.
OSIRIS has two different TFs: one for the blue range (3750--6750\,\AA; 
yet to be commissioned), and another for the red range (6510--9350\,\AA). 
In this work, we describe the results for the latter (red) etalon.

The relatively large FoV on a 10-m telescope, combined with the tunable 
nature of the filters, opens up a new method of observation, which can be 
useful in a variety of astrophysical contexts, both galactic and 
extragalactic. 
The use of narrow-band filters, adequately defined  to map certain 
spectral regions, allows the detailed study at seeing-limited resolution, of 
the spatial distribution of the  ionized gas properties and recent star 
formation, and of the associated stellar populations. Flux calibrated images 
in the bright lines of hydrogen, oxygen, nitrogen and sulphur in the optical, 
allow us to map, among other parameters, the metallicity gradients 
in galaxies using imaging techniques rather than the time-consuming 
spectroscopic techniques that are in use currently \citep{Rosa10}. 
This has motivated recently built large telescopes such as the 11-m South 
African Large Telescope \citep{Rang08} or the Magellan Baade 6.5-m telescope
\citep{Veil10} to have TFs for imaging. We are presently engaged in a survey 
of galaxies in the local universe (LUS\footnote{http://www.inaoep.mx/$\sim$gtc-lus/}= Local Universe Survey) that includes 
all galaxies inside a volume of 3.5~Mpc in radius that are visible from 
La Palma. IFU observations of these galaxies present two problems: IFU 
observations would represent  a huge drop in spatial resolution and, with a 
median D$_{25}$ of 4.5$^\prime$, they are too large to be observed even with 
the largest available IFU.

\citet{Lara10} used template spectra typical of star-forming galaxies
to estimate the errors in derived flux 
for extragalactic point-like sources on the OSIRIS TF images. However, 
photometric accuracy of the TF images of extended objects using observed data
is yet to be evaluated. In this study, we explore the capabilities of TF for 
obtaining flux calibrated images using observations acquired with OSIRIS TF.
The OSIRIS etalon is placed in a converging beam, which results in the 
variation of the effective wavelength over the FoV \citep{Born80}. 
For the red etalon, the change is as much as 80\,\AA\, over the 4$^\prime$ 
radius, thus limiting the monochromatic FoV to less than 2$^\prime$ radius
\citep{Mend11}. In addition, the TF spectral response function
(see Eq.~\ref{eqn_Rlambda}) has a long tail, which difficults the 
flux calibration of a
faint line in the neighborhood of a bright one. For example 
H$\alpha$ line contaminates even if the TF is tuned to detect the 
[\nii]$\lambda6583$ line. Given that these characteristics of TF are 
well-understood, flux calibrated monochromatic images over the entire FoV can
be obtained using a tailor-made post-observation reconstruction software.

The fact that the filters are tunable, in addition, creates its own problems for
absolute flux calibration due to the non-standard nature of the filters. 
Observing spectrophotometric standard stars at each tuned wavelength is
a solution, but it is costly in terms of telescope time. 
Alternatively, small telescopes could be used to carry out spectrophotometric 
observations of the in-frame field stars. The recent Sloan Digital Sky Survey 
(SDSS) spectrophotometric survey of field stars \citep{Beer06, Abaz09} offers 
an attractive solution to the problem, that does not require any telescope
time for calibration purposes.
In this paper, we explore the use of this database to spectrophotometrically 
calibrate the in-frame SDSS photometric stars.

In \S2, we describe the data used in this work. The method 
we followed for reconstructing a monochromatic image and its implementation
are described in \S\S3 and 4, respectively.
The sensitivity of the reconstructed
emission line fluxes to various observational parameters  is studied using simulated images in \S5 where we also 
suggest guidelines for planning imaging observations with OSIRIS/TF.
The flux calibration technique and accuracy 
are described in \S6. H$\alpha$ fluxes of selected \hii\ regions in the
reconstructed image are compared to an H$\alpha$ image obtained in a
traditional way in \S7. The results of this study are summarized in \S8.

\section{Description of the Data for Reconstruction}

\subsection{The OSIRIS detector and image formats}

OSIRIS uses two CCDs of $2048\times4102$ pixel format to cover its total 
FoV of $8^\prime\times8^\prime$, with a physical gap of 
$\sim9^{\prime\prime}$ between the two CCDs. 
Each one has 50 additional overscan pixels 
at the beginning of the array, thus resulting in a format of $2098\times4102$ 
for each image. The optical center of OSIRIS lies in the central gap 
as described in \citet{Mend11} 
and also in the OSIRIS User Manual\footnote{http://www.gtc.iac.es/en/media/documentos/OSIRIS-USER-MANUAL{\_}v1.1.pdf}.
Astrometric calibration is required before being able to mosaic the 
images registered in the two CCDs.
By performing astrometry of 10 stars on each CCD that are 
distributed over the entire FoV, we found that the image scale is slightly
different for the two CCDs. The left and right CCDs have values of 
0.127\arcsec/pixel and 0.129\arcsec/pixel, respectively, with an average
value of 0.128\arcsec/pixel. This average value agrees well with the 
theoretically expected value of the plate scale for the GTC/OSIRIS instrument 
parameters listed in Table~1 of \citet{Mend11}.

\subsection{OSIRIS TF emission-line imaging of M101}

We illustrate the reconstruction technique using observations of M101 aimed
at obtaining continuum-free images of the H$\alpha$, [\nii]$\lambda6583$, 
  [\sii]$\lambda6716$ and [\sii]$\lambda6731$ emission lines over a FoV 
of $8^\prime\times8^\prime$. The observations were carried out at 
the GTC in two observing runs on June 22 and 26 in 2009, the former for the
[\sii]$\lambda\lambda 6716, 6731$   lines and the latter for the H$\alpha$, 
[\nii]$\lambda6583$ lines. Table~1 describes the details of these two 
observing runs. The telescope pointing remained the same for all images (wavelengths) 
constituting a TF scan. The telescope position was 
then dithered by 6\arcsec\ and the entire sequence was repeated. This was done in order 
to facilitate removal of any detector artefacts. For a given region in the
galaxy, the two dithered TF images have slightly different wavelength and 
hence different response for the detection of a line. Thus it is important
to take into account these response differences before coadding them. 
We handled the dithered image sets as independent sets of data 
and used them to estimate internal errors in the flux calibration (see \S6). 
For the H$\alpha+$[\nii] scan, observations were carried out at two dithered
positions (P1 and P2, henceforth), whereas for the [\sii] scan, three 
dithered positions were used (P1, P2 and P3, henceforth).

\begin{table}
 \centering
 \caption{\label{Tab:OBs}GTC observing blocks summary}
  \begin{tabular}{lcccc}
  \hline
Observing Block  & $\lambda_{c}$ & FWHM & Order Sorter  & Exp. Time \\
\      & (\AA)        & (\AA)          & Filters & (seconds)         \\
\hline
H$\alpha+$[\nii] & 6528  & 18   & f648/28    &   $2\times180$    \\
       & 6548  & 18   & f648/28    &   $2\times180$  \\
       & 6568  & 18   & f648/28    &   $2\times180$  \\
       & 6588  & 18   & f657/35    &   $2\times180$  \\
       & 6608  & 18   & f657/35    &   $2\times180$  \\
       & 6628  & 18   & f657/35    &   $2\times180$  \\
       & 6648  & 18   & f657/35    &   $2\times180$  \\
       & 6668  & 18   & f657/35    &   $2\times180$  \\
       & 6688  & 18   &  f666/36   &   $2\times180$  \\
 \ [\sii]  & 6696  & 18   &  f666/36   &   $3\times180$  \\
       & 6716  & 18   &  f666/36   &   $3\times180$  \\
       & 6736  & 18   &  f666/36   &   $3\times180$  \\
       & 6756  & 18   &  f666/36   &   $3\times180$  \\
       & 6776  & 18   &  f680/43   &   $3\times180$  \\
       & 6796  & 18   &  f680/43   &   $3\times180$  \\
       & 6816  & 18   &  f680/43   &   $3\times180$  \\
\hline
\end{tabular}\\
\end{table}

Initial reduction of the images for correcting for BIAS was done
independently for the two CCDs constituting an image. Reliable flat-field 
frames were not available for the set of data we were analyzing, hence 
no attempts were made to correct for this. Typical pixel-to-pixel errors 
in response are expected to be less than 1\% for the CCD chips used.
Stars in the field of each of the two CCDs
were independently analyzed to obtain the astrometric solution between the
pixel and equatorial coordinates, using the SDSS coordinates of 
around 10 stars in the observed field as reference. The two CCDs were then stitched together 
to form a single image, free of all geometrical distortions. 
The procedure is repeated for the other scans. 

\section{Monochromatic Image Reconstruction Technique}

\subsection{Basic formulae for tunable filters}

The effective wavelength $\lambda_r$ at a distance $r$ from the optical 
center of the tunable filter (TF) changes following the law 
\citep{Mend11, Beck98}:
\begin{equation}
\lambda_r = {\lambda_{\rm c}\over\sqrt{1+6.5247\times10^{-9} r^2}} ,
\label{eqn_lambda}
\end{equation}
where $\lambda_{\rm c}$ is the wavelength at the optical center. The 
coefficient of $r^2$ is the inverse square of the effective focal 
length of the camera,
with $r$ expressed in physical units of pixels of 15$\mu$m in size. 
Due to the geometrical distortions of the images, a given $\lambda_r$
moves to a different radial distance $r^\prime$ in the astrometrically 
calibrated images. We used the astrometric solution to map the new 
wavelengths $\lambda_{r^\prime}$ at each astrometrically calibrated image 
with resampled pixels of 0.125\arcsec\ in size.
This correction, which is not symmetric around the optical center, 
increases quadratically with radial distance from the optical center,
reaching values of $\sim2$ and $\sim5$~\AA\ at $4^\prime$ away from the 
optical center on the left and right CCDs, respectively. 

For a TF of Full Width at Half Maximum FWHM, the filter response for a certain
wavelength $\lambda$ (e.g.~the H$\alpha$ line) in a pixel where the nominal 
wavelength is $\lambda_r$, is given by:

\begin{equation}
R_\lambda(r) = \left( 1 + \left[\frac{2(\lambda-\lambda_r)}{FWHM}\right]^2\right)^{-1}.
\label{eqn_Rlambda}
\end{equation}

This equation is an approximation valid when the
free spectral range (i.e.~the wavelength separation
between adjacent orders) is much larger than
the spectral purity (i.e.~the smallest measurable
wavelength difference) --- see, e.g., \citet{Jones02}. 
The ratio between free spectral range and spectral purity
is called finesse and is a critical parameter of all TFs.
In the case of OSIRIS, the finesse is of the order of 50
(see OSIRIS User Manual), which allows the use of Eq.~\ref{eqn_Rlambda} 
\citep{Jones02}. Spectral purity and FWHM coincide
when the finesse is large. The wavelength at the optical center
($\lambda_c$ in Eq.~\ref{eqn_lambda}) as well as the FWHM are set by tuning
the etalon. In  our observations the FWHM
was set to 18\,\AA\ (see Table 1)  rendering
a free spectral range of the order of 900\,\AA.

A monochromatic image at a particular wavelength (e.g.~H$\alpha$) over 
the entire FoV of the detector can be obtained if we have a sequence of 
images where $\lambda_{\rm c}$ between successive images increases by 
$\lesssim$FWHM. For the image sequences we have, this criterion was not
strictly met ($\Delta\lambda$ between successive images was 20~\AA\ for a 
FWHM=18~\AA). The consequence of this is that there are annular zones with 
missing data. However, multiple observations with dithered positions helped 
us to fill-in data for these zones.
In what follows we describe the method we have 
adopted for reconstructing a monochromatic image from M101 image sequences.

\subsection{Reconstruction Strategy}

The observed count rate $F(x,y)$ at a pixel $x,y$ of an image is related to 
the emitted intensity $I_\lambda(x,y)$ by the equation:

\begin{equation}
F(x,y) = \frac{\int{I_\lambda(x,y) R_\lambda(r) d\lambda}}{\kappa} + {\rm Sky}(x,y)
\label{eqn_recon1}
\end{equation}
where $R_\lambda(r)$ (Eq.~\ref{eqn_Rlambda}) is the value of the response 
curve at a distance $r=\sqrt{(x-x_0)^2+(y-y_0)^2}$ from the optical 
center ($x_0,y_0$), $\kappa$ is the 
conversion factor between the intensity and the count rate in units of 
erg\,cm$^{-2}$\,s$^{-1}$/(count\,s$^{-1}$), and $Sky(x,y)$ is the count
rate from the sky at a pixel $x,y$. In \S6, we describe a procedure 
to determine the value of $\kappa$. Ideally, the sky term can be 
estimated if observations of sky frames are taken in each scanned wavelength. 
Given that this involves extra telescope time, such observations were not 
carried out. 
The dependence of sky on $x,y$ is because of two causes (1) an intrinsic
spatial variation of the sky, and (2) a wavelength dependence of sky emission.
The first of these can be assumed to be negligible over the FoV of OSIRIS 
and hence the sky variation in the image is principally due to the second 
effect. Given the circular symmetry of the variation of wavelength, sky values 
are radially symmetric around the optical center, 
i.e. ${\rm Sky}(x,y)\equiv {\rm Sky}(r)$. 
We determined ${\rm Sky}(r)$ as the average count rate from the sky pixels in 
annular zones of width=$FWHM/2$.

In general, $I_\lambda$ includes two contributions: (1) $I_{\rm line}$, 
emission line flux integrated over the line profile, 
and (2) $I_{\rm cont}$, the continuum flux density. 
We first describe the method for reconstructing monochromatic images
in the simplest case of only one emission line 
entering the filters in the entire scanned wavelength range. 
The emission lines from nebular sources are extremely narrow with
respect to the typical bandwidth of the TF, and hence the count rate from a
line depends only on the value of the response function at the wavelength 
of the line, $R_{\rm line}(r)$. Hence the integral can be replaced by a
simple multiplication,
i.e. $\int{I_{\rm line}(x,y) R_\lambda(r) d\lambda} = 
I_{\rm line}(x,y) R_{\rm line}(r)$, when line-width $<< FWHM$.
On the other hand, the term due to the continuum from the source is
$\frac{\int{I_{\rm cont}(x,y) R_\lambda(r) d\lambda}}{\kappa}\equiv 
{\rm Cont}(x,y)$,
where $Cont(x,y)$ is the observed count rate in the continuum image.
With these approximations Eq.~\ref{eqn_recon1} can be re-written as:

\begin{equation}
F(x,y) = \frac{I_{\rm line}(x,y) R_{\rm line}(r)}{\kappa} + {\rm Cont}(x,y) + {\rm Sky}(r). 
\label{eqn_recon2}
\end{equation}

A trivial manipulation of the above equation gives:

\begin{equation}
I_{\rm line}(x,y) = \frac{\kappa \left(F(x,y) - {\rm Cont}(x,y) - {\rm Sky}(r)\right)}{R_{\rm line}(r)}
\label{eqn_recon3}
\end{equation}

By making a simple substitution 
\begin{equation}
C(x,y) = F(x,y) - {\rm Cont}(x,y) - {\rm Sky}(r),
\label{eqn_recon5}
\end{equation}
where $C(x,y)$ is the sky and continuum subtracted count rate at a 
position $x,y$ of the image, the above equation can be re-written as:

\begin{equation}
I_{\rm line}(x,y) = \frac{\kappa\ C(x,y)}{R_{\rm line}(r)}. 
\label{eqn_recon4}
\end{equation}

Due to the long tail of the response curves, an emission line from some
regions of the galaxy is registered by more than one image of the scan. 
Hence, in general,
simultaneous equations of the kind of Eq.~\ref{eqn_recon4} can be written
for each image. The line contribution in the two adjacent images
can be combined, with the weights for combining being determined 
by the values of the response functions for the line at each pixel. 
A generalized equation for recovering the flux from any pixel where the 
line is registered by two consecutive images of a scan: $im1$
and $im2$, can be written as:

\begin{equation}
I_{\rm line}(x,y) = {\left[\kappa C(x,y)\right]_{im1} + \left[\kappa C(x,y)\right]_{im2}\over{\left[R_{\rm line}(r)\right]_{im1}+\left[R_{\rm line}(r)\right]_{im2}}}
\label{eqn_recon6}
\end{equation}

For the sake of compactness, from now onwards $R_\lambda(r)$ will be denoted 
simply by $R_\lambda$. 

\subsection{H$\alpha$ and [\sii] scans}

Due to the long tail of the TF response curve, an emission line separated 
from the target line by $\sim FWHM$ contributes non-negligibly to the flux 
received in the image. This is the case while mapping the H$\alpha$ line 
with a tunable filter of FWHM$\sim18$\,\AA, where the contribution of the 
flanking [\nii] lines to the observed flux cannot be neglected. 
The contribution from an unwanted line becomes even more important
while trying to recover the [\nii] lines, as the H$\alpha$ line is likely to 
dominate the observed flux, rather than the [\nii] lines.
Another interesting case is that of the [\sii] doublet,
where both lines contribute in most of the images 
of a typical TF scan. These cases can be easily treated by adding more 
terms --- one term such as $I_{\rm line}(x,y) R_{\rm line}(r)$ for each 
emission line --- to the numerator of the first term in Eq.~\ref{eqn_recon2}.
For the scan involving the H$\alpha$ line, the first term in the 
{\it numerator} of Eq.~\ref{eqn_recon2} should be replaced by:

\begin{equation}
I_{\rm line}(x,y) R_{\rm line} \rightarrow
I_{{\rm H}\alpha} R_{\lambda{\rm H}\alpha} 
 + I_{6583} R_{\lambda 6583} 
 + I_{6548} R_{\lambda 6548}, 
\label{eqn_recon7}
\end{equation}
where $R_{\lambda{\rm H}\alpha}$, $R_{\lambda 6583}$, and
$R_{\lambda 6548}$ are the values of the response function 
at a radial distance $r$ from the optical center
for the observed (not the rest-frame) wavelengths of the corresponding lines.

Eq.~\ref{eqn_recon4} for the ${\rm H}\alpha$ line can then be re-written as,

\begin{equation}
I_{{\rm H}\alpha}(x,y) = \frac{\kappa C(x,y)}{R_{\lambda{\rm H}\alpha}+\frac{I_{6583}}{I_{{\rm H}\alpha}}\left(R_{\lambda 6583} + \frac{1}{3} R_{\lambda 6548}\right)}, 
\label{eqn_recon8}
\end{equation}
where we have substituted the value of the intrinsic flux ratio 
of $\frac{I_{6548}}{I_{6583}}=\frac{1}{3}$. The equation has a term
involving $\frac{I_{6583}}{I_{{\rm H}\alpha}}$ in the denominator.
Thus in order to reconstruct the H$\alpha$ line image, we need to know 
{\it a priori} the flux ratio images of the contaminating [\nii]$\lambda6583$ 
line with respect to the H$\alpha$ line. 
The [\nii] line fluxes, on the other hand, require a knowledge of 
the H$\alpha$ flux, as can be seen by the equation for the 
[\nii]$\lambda6583$ line:
\begin{equation}
I_{6583}(x,y) = \frac{\kappa C(x,y) - {I_{{\rm H}\alpha}}(x,y) R_{\lambda{\rm H}\alpha}}{R_{\lambda 6583}}. 
\label{eqn_recon9}
\end{equation}
In this equation, we have neglected the [\nii]$\lambda6548$ line 
contribution given that it contributes less than $2\%$ in image sections 
tuned to maximize the [\nii]$\lambda6583$ emission.
Thus, in order to obtain [\nii]$\lambda6583$ flux map,
we need to know the H$\alpha$ flux. An iterative procedure involving 
Eqs.~\ref{eqn_recon8} and ~\ref{eqn_recon9} is necessary for an accurate 
recovery of both H$\alpha$ flux and $\frac{I_{6583}}{I_{{\rm H}\alpha}}$ ratio.
Given the exploratory nature of the present study, we have calculated the
H$\alpha$ fluxes by fixing $\frac{I_{6583}}{I_{{\rm H}\alpha}}=0.1$ 
for all regions. This approximation would result in an overestimation of 
the H$\alpha$ fluxes by $>2.5$\%, and underestimation
of the $\frac{I_{6583}}{I_{{\rm H}\alpha}}$ ratios by $\sim0.1$
for regions having intrinsic $\frac{I_{6583}}{I_{{\rm H}\alpha}}\gtrsim0.2$.

The reconstruction equations for the $\lambda6716$ and $\lambda6731$ lines 
from the [\sii] scan are:

\begin{equation}
I_{6716}(x,y) = \frac{\left[\kappa C(x,y)\right]_{im1} + \left[\kappa C(x,y)\right]_{im2}}
{R^{im1}_{\lambda 6716} +R^{im2}_{\lambda 6716} +\frac{I_{6731}(x,y)}{I_{6716}(x,y)} (R^{im1}_{\lambda 6731}+R^{im2}_{\lambda 6731})}, 
\label{eqn_recon10}
\end{equation}

and

\begin{equation}
I_{6731}(x,y) = \frac{\left[\kappa C(x,y)\right]_{im1} + \left[\kappa C(x,y)\right]_{im2}}
{R^{im1}_{\lambda 6731} +R^{im2}_{\lambda 6731} +\frac{I_{6716}(x,y)}{I_{6731}(x,y)} (R^{im1}_{\lambda 6716}+R^{im2}_{\lambda 6716})}, 
\label{eqn_recon11}
\end{equation}
In these equations, $im1$ and $im2$ denote the image sections tuned 
to maximize the $\lambda6716$ and $\lambda6731$ lines, respectively.
Note that, in order to recover the $I_{6716}$ image, we need the $I_{6731}$ 
image and vice versa. Unlike the case of 
$\frac{I_{\rm N\,II}}{I_{{\rm H}\alpha}}$, no apriori values for 
the intensity ratios of the [\sii] lines could be used, given that this ratio
is very sensitive to the electron density of the regions.
We hence resolved Eqs.~\ref{eqn_recon10} and \ref{eqn_recon11}
by assuming a value of 
$\frac{I_{6716}}{I_{6731}}$, and iteratively changing that value until the 
value at every pixel stabilizes within 10\% in two successive iterations. 
We checked that the resulting images
are the same irrespective of the starting value of $\frac{I_{6716}}{I_{6731}}$.

\section{Implementation of the method}

\begin{figure}
\begin{center}
\includegraphics[scale=.5]{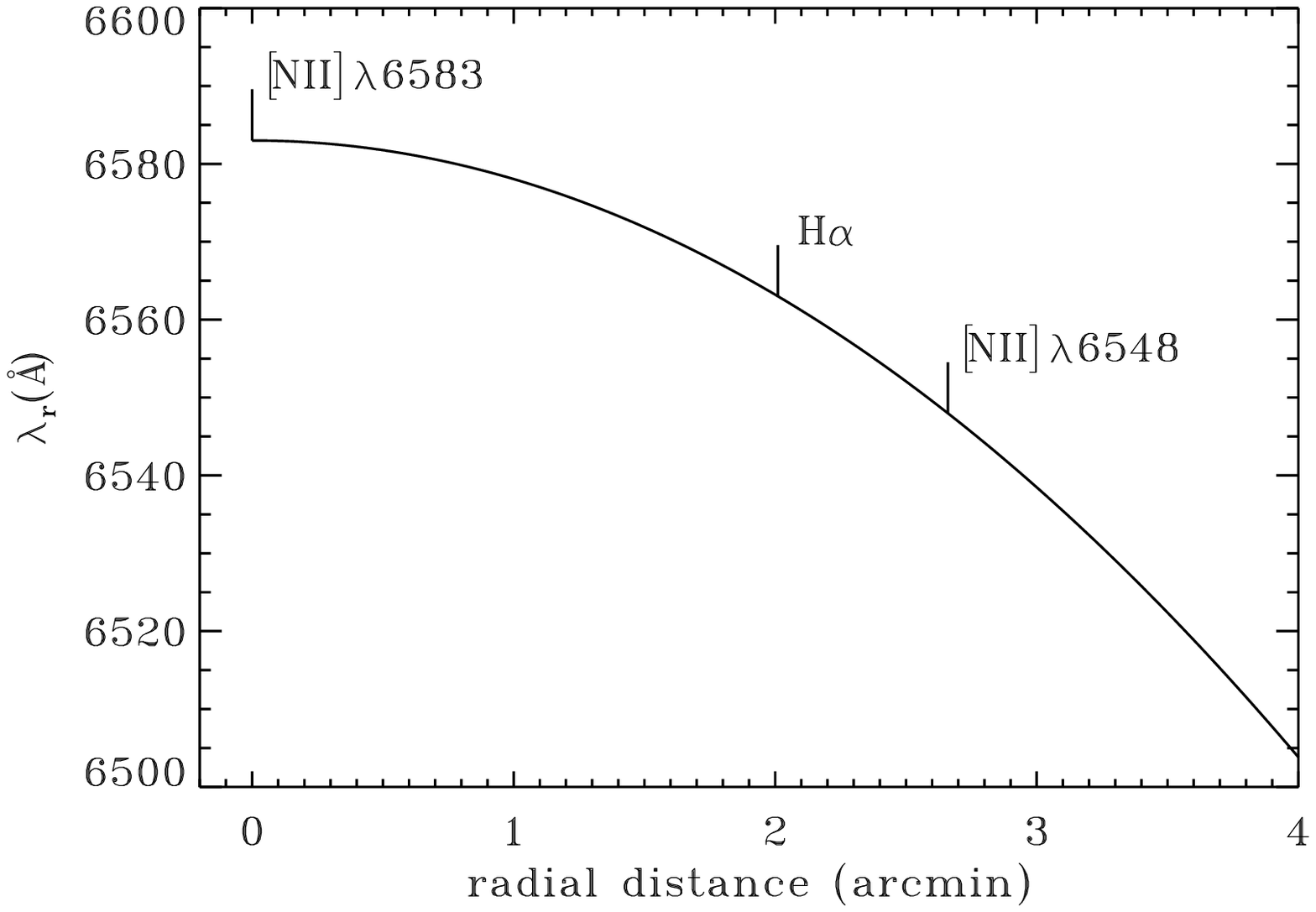}
\includegraphics[scale=.5]{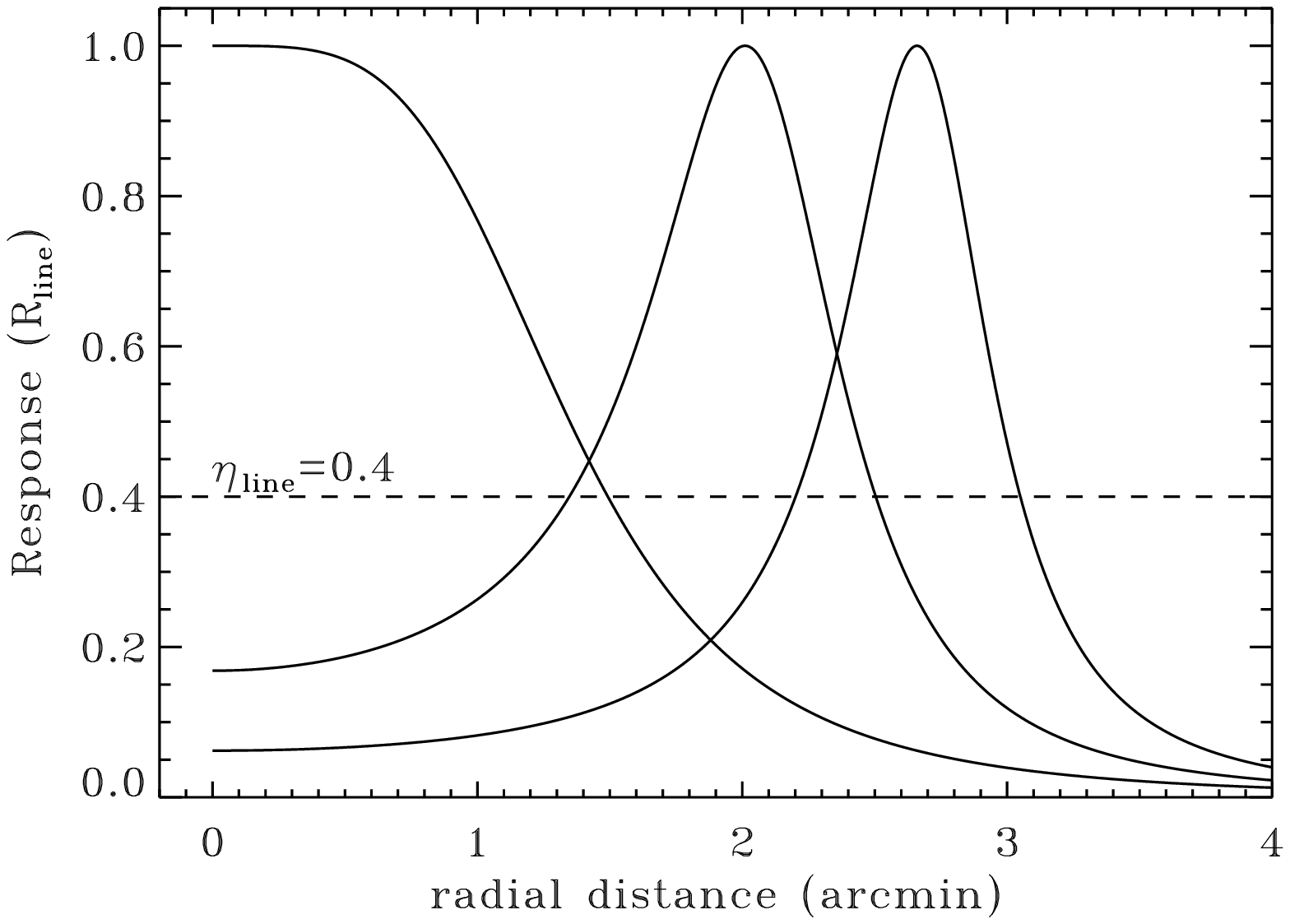}
\caption{
(Top) An illustration of the variation of the wavelength over the OSIRIS FoV
as expected from Eq.~\ref{eqn_lambda}, with $\lambda_{\rm c}=6588$~\AA\  
(observed wavelength of [\nii]$\lambda6583$ line in M101) and 
FWHM=18~\AA. The positions where [\nii]$\lambda6583$, H$\alpha$,
[\nii]$\lambda6548$, lines have maximum response are marked.
(Bottom) The response curves, as calculated using Eq.~\ref{eqn_Rlambda}, 
for each of these lines are shown. The part of the curves above the 
horizontal line ($\eta_{\rm line}=0.4$) can be used to reconstruct the 
monochromatic images in the corresponding lines.
\label{fig_lambda}}
\end{center}
\end{figure}

We developed a script under the IRAF\footnote{IRAF is distributed by the National Optical 
Astronomy Observatory, which is operated by the Association 
of Universities for Research in Astronomy (AURA) under 
cooperative agreement with the National Science Foundation.}  environment to implement the method
in a user-friendly way. The first step of the reconstruction process is 
obtaining wavelength {\it vs.} pixel look-up images for every image
of the scan. These images are created using Eq.~\ref{eqn_lambda} and 
the astrometric solutions
as described in \S3.1. Values of $\lambda_{\rm c}$\footnote{
Wavelength calibration lamps were not supplied in this initial observing run, 
and hence we checked/recalibrated the value of $\lambda_{\rm c}$ by comparing 
the filter-convolved SDSS spectra of 7 \hii\ regions to the profiles of the 
corresponding \hii\ regions formed using the observed fluxes in successive 
images. Error in $\lambda_{\rm c}$ using this method is found to be 
$\lesssim2$\AA, which is better than what could be achieved using the sky rings. 
} 
and FWHM are taken from Table~1.
For each of the emission lines, we then created 2-dimensional response 
images using Eq.~\ref{eqn_Rlambda}. The procedure we followed is 
illustrated using 1-d cuts on one of these images in Fig.~\ref{fig_lambda}.  
Specifically, we chose the image with $\lambda_{\rm c}=6588$~\AA\ for
illustration, as this image is capable of detecting three different 
emission lines at different radial bins. The bottom panel
shows the expected value of the response functions for the three 
lines as a function of the radial distance from the optical center.
In cases such as this, where more than one emission line is registered within
the $8^\prime\times8^\prime$ FoV ($r=4^\prime$), a second line starts 
contributing significantly before the line of interest drops to below 
$\sim$50\% response, as illustrated in the bottom panel of
Fig.~\ref{fig_lambda}.
It is possible to isolate the contribution of each line 
by selecting only those zones where the line of interest has a response
 value above a certain level, as described below.

\subsection{Choosing the value for the response cut-off}

The part of a TF image that can be considered monochromatic is decided by the 
value for the response cut-off ($\eta_{\rm line}$) parameter. 
By carefully choosing $\eta_{\rm line}$, it is possible to reconstruct
a monochromatic image in the line of interest even in the presence of 
contaminating lines.
Only those pixels with a response value $R_{\rm line}>\eta_{\rm line}$ 
will be considered good for reconstructing the monochromatic image in
that $line$. All these pixels belong to an annular zone of particular width 
(circle in the case of a line falling close to the optical center).
For example, with $\eta_{\rm line}=0.4$, the image sections for reconstructing 
the H$\alpha$ and [\nii]$\lambda6583$ lines in the $\lambda_{\rm c}=6588$~\AA\ 
image correspond to the radial zones where the curve lies
above the horizontal line in Fig.~\ref{fig_lambda}. 

Two guidelines are useful for setting the value of $\eta_{\rm line}$:
(1) the response for the contaminating line in the selected image section
has to be less than that for the main line, 
(2) there are no annular gaps
in the final reconstructed image. The first condition depends on 
the wavelength difference between the contaminating lines, whereas the 
difference between central wavelengths of successive images of the scan 
($\Delta\lambda_{\rm c}$)
determines the fulfillment of the second condition.
For the case of H$\alpha$ and [\nii]$\lambda6583$, the response value  
would be the same at a wavelength mid-way between the two lines
(i.e. 6573~\AA\ at rest-frame or 6578~\AA\ for our M101 scan given its 
recession velocity of 214 km\,s$^{-1}$). At radial zones where 
$\lambda_r<6578$~\AA, the 
observed count rates would be predominantly from the H$\alpha$ line,
implying $\eta_{\rm line}\geq0.45$.
On the other hand, the second criterion requires that $\lambda_r$ be separated from
$\lambda_{{\rm H}\alpha}$ by at least $\Delta\lambda_{\rm c}/2$, implying
$\eta_{\rm line} \leq 0.45$ for our M101 scan. Thus $\eta_{\rm line}=0.45$ is the
optimal value for our dataset. The value of $\eta_{\rm line}$ could be
marginally higher than this if dithered images are able to fill in data-less annular 
zones, or lower if the contaminating line contribution could be subtracted 
to within a few percent accuracy using an iterative procedure. 
For example, for our M101 scan, dithering between the images  
was sufficient to be able to fill-in the data gaps for $\eta_{\rm line}=0.5$.
On the other hand, the SNR of the [\nii]$\lambda6583$ images was not good
enough to iteratively subtract the [\nii]$\lambda6583$ contamination in the
predominantly H$\alpha$ pixels. The user would be required to select the
value of $\eta_{\rm line}$ based on the data parameters and the specific
scientific objective.

\subsection{Coadding monochromatic image sections}

\begin{figure}
\begin{center}
\includegraphics[scale=.50]{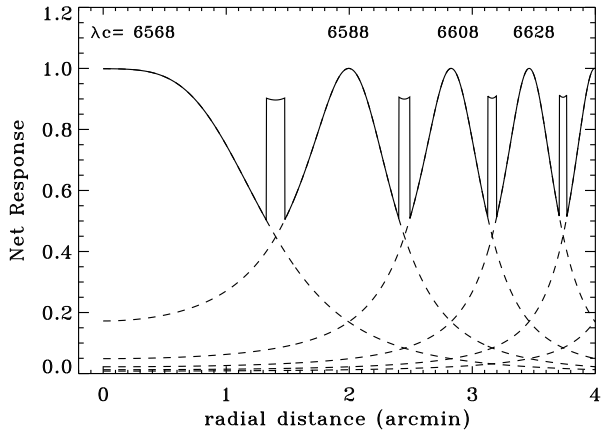}
\caption{
The net response curve for the reconstruction of the H$\alpha$ image
for $\eta_{\rm line}=0.4$ is shown by the solid curve.
H$\alpha$ response curves in individual TF images are shown by the 
dashed lines. The central wavelength $\lambda_{\rm c}$ (in \AA) for
each image is indicated above the corresponding response function.
\label{fig_netresp}}
\end{center}
\end{figure}

In Fig.~\ref{fig_netresp}, we show the response for the
H$\alpha$ line in consecutive images of a TF scan. For a value 
of $\eta_{\rm line}=0.4$, in certain ranges of radial zones, the line 
is registered in two consecutive images. Thus, for $\eta_{\rm line}=0.4$
there is redundant data for some pixels 
for the construction of the monochromatic images. This redundancy can be 
used to our advantage to get deeper images, by coadding the pixel values
from both images that contribute to that zone. The response curves
are also coadded to get a net response curve such as shown by the
solid line in Fig.~\ref{fig_netresp}. The coadded line image is
divided by the net response image to get the entire image in the same
flux scale (see Eq.~\ref{eqn_recon6}).

\subsection{Preparation of the continuum and sky images}

Fundamental to the reconstruction process is the determination of pure
count rate in the line $C(x,y)$ from the observed count 
rate $F(x,y)$. This is carried out through Eq.~\ref{eqn_recon5}, 
which involves the estimation of sky and continuum images. The sky count 
rates $Sky(r)$ are calculated as the median values in semi-annular zones, 
with a width equal to $FWHM/2=9$~\AA. 

The bluest wavelength of our scans is $\lambda_c=6528$~\AA, which is the 
filter with least contamination from any emission line. We used this image 
as a first-guess image of the continuum. Even for this filter, the H$\alpha$ 
line contributes more than 3\% for pixels closer than 90\arcsec\ from the 
optical center. We defined a parameter $\eta_{\rm cont}$ such that only those 
pixels for which the response to detect any line does not exceed 
$\eta_{\rm cont}$ are considered for continuum (i.e.~only pixels with 
$R_{line}<\eta_{\rm cont}$). For example, for the image 
with $\lambda_c=6528$\AA, with $\eta_{\rm cont}=0.03$ only pixels that 
are more distant than 90\arcsec\ from the optical center satisfy this 
condition. Even for these pixels, we estimated the line contribution using 
the reconstructed emission line maps, and subtracted it from the continuum 
pixels. Individual sections contributing to the continuum are stitched 
together (averaged when more than one image contributes to a pixel) to obtain
the final continuum image. This image is input as the new continuum image
and the entire process is repeated. From the simulated data (see \S5.3, {\it 
Run 1}), we found that the continuum values converge after 3 iterations. 
A small value of $\eta_{\rm cont}$ would result in various annular zones 
where there are no pixels satisfying the 
condition $R_{\rm line}<\eta_{\rm cont}$.
From simulated data, we found that line-free continuum images could be
obtained even for $\eta_{\rm cont}$ as large as 0.2.

An IRAF package containing the scripts developed as part of this work is 
available to users upon request and will be available for downloading 
from the LUS webpage http://www.inaoep.mx/$\sim$gtc-lus/.

\section{Flux errors in the reconstructed images}

\subsection{Flux error due to tuning error}

The central wavelength of the OSIRIS TF can be set only to an accuracy
of 1\,\AA\ (See OSIRIS TF User Manual). This uncertainty in tuning the TF
leads to an error in the recovered emission line fluxes.
The error ($\delta F$) in flux ($F$) due to an uncertainty of 
$\delta\lambda_{\rm c}$ in setting the central wavelength 
depends directly on the first derivative of the response function 
(Eq.~\ref{eqn_Rlambda}) with respect to $\lambda$. 
\begin{equation}
i.e. \hspace{2cm} {\delta F\over{F}} \equiv {\delta R_\lambda\over{R_\lambda}} 
\equiv {1\over{R_\lambda}}\left({\partial R_\lambda\over{\partial\lambda}}\right)\delta\lambda_{\rm c}  
\label{eqn_dfbyf1}
\end{equation}
Substituting the value of ${\partial R_\lambda\over{\partial\lambda}}$, we get
\begin{equation}
{\delta F\over{F}} = 4 R_\lambda \left({\delta\lambda_{\rm c} \over{FWHM}}\right) \sqrt{\left({1\over{R_\lambda}} - 1\right)}. 
\label{eqn_dfbyf2}
\end{equation}
Maximum error is introduced for image pixels where the TF response $R_\lambda$
for the line of interest is 0.5. Thus, flux errors in the reconstructed image
can be reduced if we use $\eta_{\rm line}>0.5$.
It may be recalled from the analysis in \S4.1 that the lower limit of
$\eta_{\rm line}$ depends on $\Delta\lambda_{\rm c}$. In order to have 
$\eta_{\rm line}>0.5$, $\Delta\lambda_{\rm c}$ should be less than the $FWHM$. 

\begin{figure}
\begin{center}
\includegraphics[scale=.50]{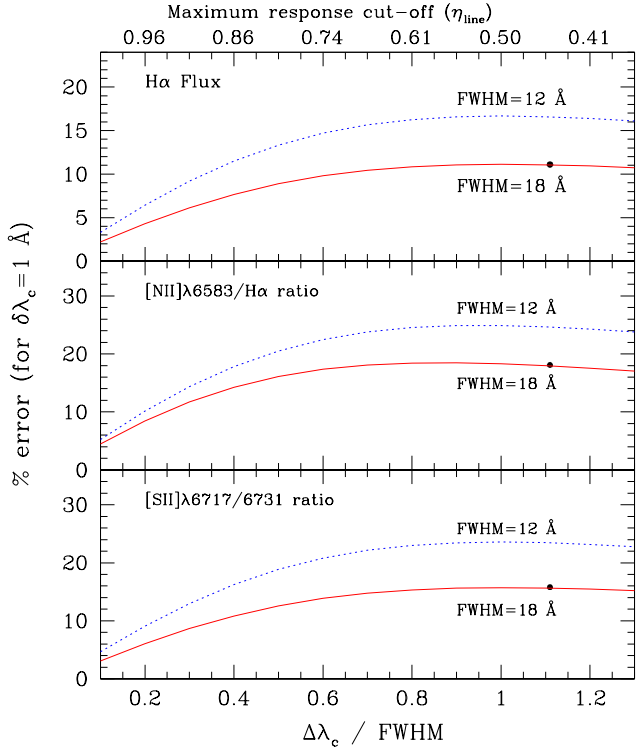}
\caption{
Errors in flux and flux ratios due to an uncertainty of 
$\delta\lambda_{\rm c}=1$~\AA\ in setting the central wavelength of TF, as 
a function of user-selected TF parameters $\Delta\lambda_{\rm c}$ and FWHM. 
The errors are expressed as percentage values of the plotted quantity. 
(top) Errors on the H$\alpha$ flux, (middle) errors on 
the [\nii]$\lambda6583$/H$\alpha$ ratio, and (bottom) errors on 
the [\sii]$\lambda6717/6731$ ratio. The scale on the top gives the maximum 
value of response cut-off ($\eta_{\rm line}$) that the data permit for given 
values of $\Delta\lambda_{\rm c}$ and FWHM, without having data gaps in the 
reconstructed image (see \S4.1 for details). Errors expected in our dataset 
of M101 for an assumed error of $\delta\lambda_{\rm c}=1$~\AA\ are marked 
by solid circles. 
\label{figures_simul}}
\end{center}
\end{figure}

We used the Eq.~\ref{eqn_dfbyf2} to calculate the errors in the line 
fluxes, and summed the errors in quadrature to calculate the errors in the 
flux ratios, for nominal value of $\delta\lambda_{\rm c}=1$~\AA. 
Our results are summarized in Fig.~\ref{figures_simul}. 
In the three panels, we plot the errors in the calculated H$\alpha$ fluxes 
(top), the [\nii]$\lambda6583$/H$\alpha$ ratio (middle) 
and [\sii]$\lambda6717/6731$ ratio (bottom) as a function of the
sampling parameter, for two extreme values of FWHM permitted by the TF.
The value of $\eta_{\rm line}$ is calculated in such a way that there are 
neither gaps nor overlapping pixels in the reconstructed image for a given 
sampling $\Delta\lambda_{\rm c}$. 

The most notable characteristic in these plots is that the
errors in all these three quantities are lower for larger values of FWHM.
This may seem counter-intuitive, and is the result of a sampling error
of 1~\AA\ being a larger fraction of $FWHM=12$~\AA\ than that for $FWHM=18$~\AA.
In other words, the TF response function falls less steeply for larger FWHM, 
thus resulting in little errors as compared to that for smaller FWHMs. 
As expected, finer sampling results in smaller errors on the derived 
quantities, especially when the sampling rate is less than $0.7\times FWHM$. 

Line fluxes can be calculated with better accuracy than the flux ratios.
The error in the [\nii]$\lambda6583$/H$\alpha$ flux ratio takes into account 
the contamination by H$\alpha$ in the [\nii]$\lambda6583$ filters. This cross 
talk also makes the errors on the [\nii]$\lambda6583$ line flux marginally
larger than those for the H$\alpha$ flux. 
The calculated errors on the individual [\sii]  line fluxes are similar to that
for the H$\alpha$ flux. 

It may be noted that a recessional velocity of 45\,km\,s$^{-1}$ produces a 
shift of 1\AA\ in the wavelength. Thus, in regions where kinematic deviations
of this order or more are likely to exist (e.g. nuclear regions of galaxies), 
kinematical data are required for an accurate flux calibration. In the 
absence of such data, eq.~15 can be used, where $\delta\lambda_{\rm c}$ is to 
be replaced by the uncertainty in the wavelength due to Dopper effect, to 
estimate the flux errors due to kinematic deviations.

\subsection{Reconstruction of simulated TF images}

The dataset we have for M101 is reconstructed with $\eta_{\rm line}=0.5$,
implying maximum errors of around 11\% in the recovered H$\alpha$ fluxes
according to Fig.~\ref{figures_simul}. In order to check the implementation
of the equations of \S3 in our reconstruction scripts, we need datasets 
with much smaller errors. We hence 
created artificial datasets by simulating the observations of an extended 
emission line source with a scanning TF, making use of Eqs. 1 and 2. 
The simulated dataset also allowed us 
to quantify the accuracy of reconstruction 
to various instrumental parameters that have non-zero uncertainties.
Two scans were simulated:
(1) TF with a FWHM=18~\AA, $\lambda_c=6528+(i-1)\times20$~\AA,
where $i$ varied from 1 to 9, 
(2) TF with a FWHM=12~\AA, $\lambda_c=6528+(i-1)\times10$~\AA,
where $i$ varied from 1 to 18. 
The reddest wavelength in both cases is 6688~\AA. 
The FWHM in the two settings are among the 
extreme values permitted with the red etalon, with the former one simulating
the dataset we have for M101. The intensity of the extended source 
is set constant over the FoV of OSIRIS with I(H$\alpha$)=1.
The [\nii]$\lambda$6583 line is fixed at 10\% of that of H$\alpha$, which
is among the range of values observed in \hii\ regions \citep{Deni02}.
The [\nii]$\lambda$6548 line intensity is fixed at one third of that of the 
[\nii]$\lambda$6583 line. We also added continuum sources at some fixed 
positions on the image, in order to test the accuracy of the continuum 
subtraction. The intensities of the continuum sources are adjusted such that 
the H$\alpha$ emission equivalent widths at the positions of the continuum 
sources are 10 for some sources and 100 for others. 
Gaussian noise of two different values is added to the images, resulting in
two sets of simulations:
the high and low Signal-to-Noise Ratio (SNR) images correspond to SNRs of 
100 and 3 for the
[\nii]$\lambda$6583 line image, respectively. A line-free continuum image is 
also generated in the simulated set. The lines were redshifted by 
214 km\,s$^{-1}$, corresponding to the recession velocity of M101.

\subsection{Parameters that limit the reconstruction accuracy}

{\it Reconstruction with ideal datasets (Run 0):}
The equations described in \S3 should allow us to recover the fluxes of the
H$\alpha$ and [\nii]$\lambda$6583 lines, in the absence of 
uncertainties in various parameters that control the TF imaging.
We simulated this case by using high 
SNR data and subtracting the simulated line-free continuum.
The errors in recovering the H$\alpha$ flux and the flux ratio 
[\nii]$\lambda$6583/H$\alpha$ are as small as  0.03\% and 0.2\% respectively.
This test establishes that the 
equations developed in \S3 are correctly implemented in the script.

However, there are many sources of error in real observational datasets,
especially when trying to maximize the available observing time. 
We investigate the accuracy in the recovered flux due to the following 3
sources of error:
(1) the continuum images are not completely free of emission lines,
(2) the sky images are obtained by in-frame object-free pixels, not from an
off-source sky image,
(3) different images within a scan have non-zero ditherings.
We discuss each of these cases in detail below. 

{\it Line contamination in continuum images (Run 1):} 
Due to the long tail of the TF response, there is non-zero line 
contribution in all TF images, including those centered bluer than the
bluest line in a scan (e.g., $\lambda_{\rm c}=6528$ is only $\sim1$ FWHM 
blueward of the [\nii]$\lambda6548$ line, and $<2$ FWHMs of the bright 
H$\alpha$ line).
In order to obtain line-free continuum fluxes from the sequence of images
we have, we followed the iterative procedure described in \S4.3. We obtained
a continuum image from the sequence of simulated images, and compared it to
the simulated pure continuum image. The difference between the two images is
found to be as small as 0.1\% after 3 iterations. Hence, errors
on the fluxes in our reconstructed images of M101 are not dominated by the 
limitations of obtaining line-free continuum images.

{\it Non-uniform sky (Run 2):} 
Errors in sky value subtraction and flat-field corrections introduce
residuals in the sky value locally. We parametrized this error in terms of the
rms noise value ($\sigma$) of this image. The residual sky value in 
different parts of the image is allowed to vary between $-1\sigma$ and 
$+1\sigma$, for the simulated set of images.
The [\nii]$\lambda$6583/H$\alpha$ ratios are affected by 5\%
in pixels where [\nii]$\lambda$6583 is detected with a SNR of $3$. 
Thus, for pixels with SNR$>3$ for the line of [\nii]$\lambda$6583, 
sky subtraction is not a serious problem in our images of M101.

{\it Consequences of Image-dithering (Run 3):} 
It is a normal practice to dither images between any two repeat
observations in order to avoid detector blemishes spoiling any interesting
feature. These dithered images are registered to a common coordinate system
before combining them. In a TF observation, the optical centers in the 
dithered images correspond to different astrometric coordinates 
and hence the corresponding pixels in the registered images do not
have the same wavelength (see Eq.~1). Thus, in the combined image,
a given pixel has contributions from marginally different wavelengths.
We parametrized this effect as an error in the optical center 
$\delta r_{\rm c}$. We studied the reconstruction accuracy for 
various values of $\delta r_{\rm c}$ between $1\arcsec$ and $5\arcsec$. 
A dithering of $1\arcsec$ 
between different images of a scan can produce errors of the order of
$\sim5$\% in the recovered H$\alpha$ flux for the FWHM=18~\AA\ filters. 
The recovered [\nii]$\lambda$6583/H$\alpha$ ratio lies between 
0.05--0.15 (50\% error over the simulated value of 0.1)
for this case. Combining images that are dithered by more than
$1\arcsec$ would introduce more than 20\% error on the recovered H$\alpha$
flux, making it unusable for most applications.
In the case of our observations of M101, the dithering within a scan
was less than $0.25\arcsec$, and hence the error in the recovered
H$\alpha$ flux due to astrometric registration of all images of a scan 
(say P1) is less than a few percent.

\subsection{Recommendations for observing extended sources}

After studying the effect of various parameters on the fluxes in the 
reconstructed images, we find that the maximum error on our dataset for M101 
arises due to the present uncertainty of $\sim1$~\AA\ in setting the central 
wavelength of a TF observation. This uncertainty affects more the images
taken with narrower TF observations, with the accuracy of H$\alpha$ fluxes 
being $\sim$10\% and of [\nii]$\lambda$6583/H$\alpha$ ratios of $\sim18$\%  
for the TF images with $FWHM=18$~\AA, as can be inferred from 
Fig.~\ref{figures_simul}. Corresponding errors with $FWHM=12$~\AA\ are 16\% 
and 25\%, respectively. Errors may be reduced by carrying out the scan with a 
finer sampling. With $FWHM=12$~\AA\, a sampling of around 5~\AA\ would be 
required to reduce the error levels to $\sim10$\%, that can be achieved 
with 15~\AA\ sampling with $FWHM=18$~\AA. Thus a factor of 3  more exposure 
time would be required to map a given emission line over the entire FoV of 
OSIRIS with the narrower TF. However, for low SNR pixels or regions for 
which sky and/or continuum subtraction errors contribute more than the 
plotted errors ($SNR\lesssim 10$), the errors are 
expected to increase proportionally with the FWHM, and hence observations with 
$FWHM=12$~\AA\ would have an advantage by a factor of 1.5 over those with 
$FWHM=18$~\AA. Thus, if the interest is in detecting diffuse faint emissions,
a $FWHM=12$~\AA\ is preferable.

\citet{Lara10} carried out simulations to determine the best combination of 
FWHM and sampling ($\Delta\lambda_{\rm c}$) for optimal emission line flux
determinations of emission-line galaxies of redshifts between 0.2 and 0.4 with
GTC/OSIRIS. They found that a FWHM of 12~\AA\ and a sampling of 5~\AA\ are 
the optimal combination that allows deblending H$\alpha$ from 
the [\nii]$\lambda6583$ line with a flux error lower than 20\%. It is 
relevant to note that in their simulations, the fluxes were not
corrected for the response curve of the TF, and the quoted flux errors
are due to the unavailability of the redshifts of the detected
galaxies. Hence, it is natural to expect lesser errors for the narrower 
filters.

The next source of error comes from image dithering. For an image dithering
of 1\arcsec, the errors could be as large as 5\% for H$\alpha$ and 50\%
for [\nii]$\lambda$6583/H$\alpha$. For larger values of dithering, errors
on [\nii]$\lambda$6583/H$\alpha$ are unreasonably high.
It is advisable to have the same telescope position (dithering $<1$\arcsec) 
for all the images constituting a given TF scan. If more than one scan is 
available for a field with dithering of  more than 1\arcsec\ between the scans,
it is advisable to reconstruct an emission-line image from each
scan, and then register and combine them. We recommend that a TF scan
intended to obtain a monochromatic image in an emission line should have
one image observed at least one $FWHM$ blueward of the bluest line in the
TF sequence, to facilitate accurate continuum subtraction.

\section{Flux Calibration of Monochromatic Images}

\begin{figure*}
\begin{center}
\centerline{\epsfig{file=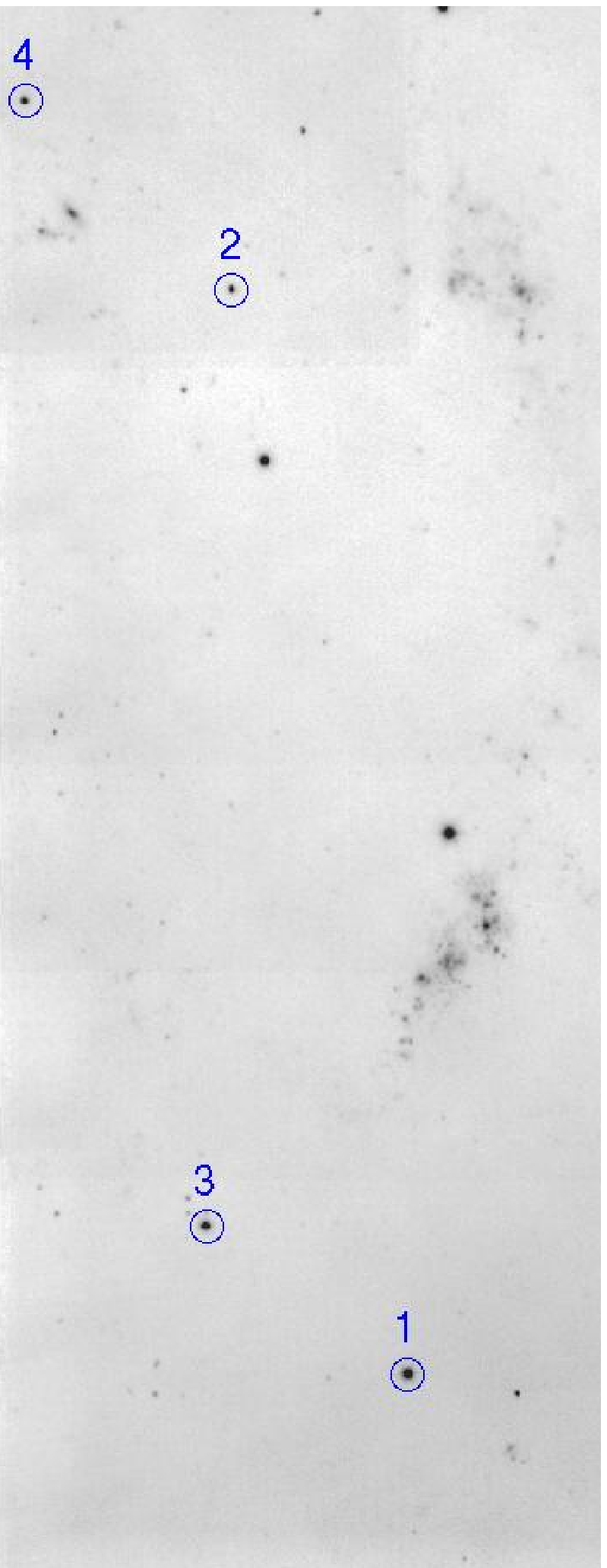, height=0.4\vsize}
            \epsfig{file=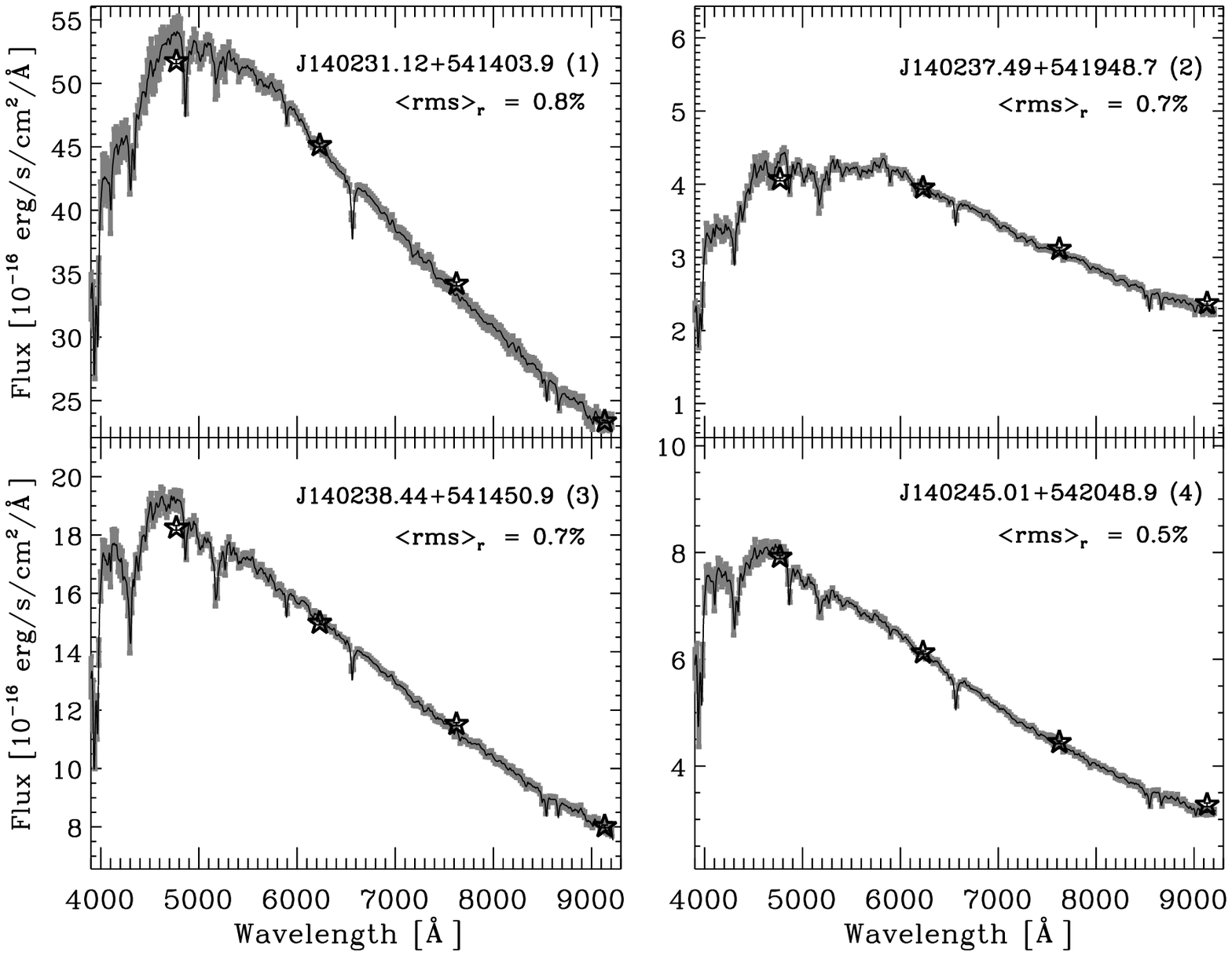, height=0.4\vsize}}
\vspace{1cm}
\caption{
An illustration of the calibration procedure adopted, where we fit the
SDSS $griz$ photometry of in-frame field stars with the spectra of 
stars in the SDSS spectral catalog. (left) A section of the
OSIRIS continuum image of M101. Four stars with SDSS photometry are identified. 
(right) SEDs in $griz$ bands of the 4 identified stars (asterisk), superposed on the 
average of 12 best fitting spectra from SDSS spectral catalog (line).
The rms deviation from the mean of 12 best-fit spectra at each wavelength is
shown by the gray band around the mean spectrum. The 12 spectra differ by 
less than 1\% ($<rms>_r$ denotes the average rms value within the $r$-band),
which makes this method very attractive for the calibration of OSIRIS TF data.
\label{fig_sdss}}
\end{center}
\end{figure*}

We have explored a new procedure to flux calibrate the TF images using the 
in-frame field stars. The procedure, in principle, can be applied to any 
optical narrow-band image of a field that contains stars with photometry
in the optical broad-bands. In particular, we used the SDSS photometry 
in $griz$ bands of stars in the field of M101, and the recently released SDSS Stellar Spectral 
database of stars covering a wide range of spectral types \citep{Abaz09}.
These spectra are of median resolution in the 3500-9000\,\AA\ wavelength range. 
In order to have a good spectral library we selected, from the original Stellar 
Spectral database, only those stars with $g=$14--18~mag having no gaps, 
jumps or emission lines in their spectra.

The general procedure involves obtaining a stellar spectrum that fits 
the $griz$ magnitudes of the stars in the field of the TF image.
Basically, we are using the stellar spectrum to interpolate the broad-band
fluxes at the wavelength of the TF. The spectra are integrated in 
the $griz$-bands to get their synthetic magnitudes which are then fitted to 
the $griz$ magnitudes of a star in the field of interest;
both Spectral Energy Distributions (SEDs) are normalized at the $r$-band.

The best-fit spectrum is chosen by minimizing the $\chi^2$ obtained 
in the 4 bands. The spectrum of the best-fit SDSS star is then 
de-normalized by multiplying it by the $r$-band flux of the field star. 
The resulting spectrum is convolved with the response function of the OSIRIS 
TFs (Eq.~\ref{eqn_Rlambda}) to obtain a smoothed spectrum of the field 
star. The flux at $\lambda_r$, where $r$ is the radial distance of the star 
in the OSIRIS field, is multiplied by the effective bandwidth of the TF, to 
estimate the flux intercepted by the TF. 

We then carried out aperture photometry of the selected SDSS stars on all the 
TF images. This photometry is used to obtain observed count rate of each
star in each TF image.
The estimated flux is divided by the observed count rate of the star in that TF to
obtain the calibration coefficient $\kappa$. 
The procedure is repeated for all the good SDSS stars in the observed field
to obtain a set of $\kappa$ values. We note that the 
observed count rate is initially corrected for the effects of extinction and 
the efficiency of the order sorter filter, and hence the $\kappa$
obtained from different stars in different TFs can be directly compared
with each other.

\subsection{Relative errors in the Calibration coefficients}

The availability of thousands of stellar spectra covering the entire range
of spectral types to fit the photometric data of field stars ensures that 
there is at least one spectrum that truly represents the spectrum of the 
field star. Generally, there are 10--15 spectra whose $griz$-band $\chi^2$ 
is within 10\% of the best-fit spectrum.
We obtained a mean and rms of these spectra at every sampled wavelength. The 
rms error was found to be less than 1\% for wavelengths between the $g$ and $z$ bands.
Thus the relative flux calibration for different TF settings is better than 
1\%. The absolute flux error depends on the error in the SDSS $r$-band magnitude
of the field stars. An illustration of the method followed is 
shown in Fig.~\ref{fig_sdss}.

\begin{figure}
\begin{center}
\centerline{\epsfig{file=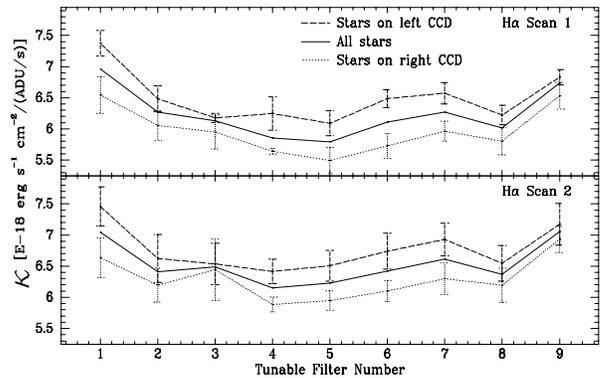, height=0.65\hsize}}
\caption{
Variation of the calibration coefficient $\kappa$ for 9 images
of the two H$\alpha$ TF scans of M101. We used 8 stars, 4 on each CCD 
covering the complete field of view. The coefficients for the right 
(dashed line) and left (dotted line) CCDs are separately shown. The solid 
line shows the variation without distinguishing on which CCD stars are located.
In the top panel, we show the coefficients for scan 1 (P1) and in the
bottom, for scan 2 (P2).
\label{fig_calha}}
\end{center}
\end{figure}

In Fig.~\ref{fig_calha}, we show the variation of the calibration 
coefficient along an H$\alpha$ TF scan containing 9 images. 
In this figure, two kinds of systematic variations are seen: 
(1) the variation of the coefficient from one TF observation to the next,
and (2) the variation of the coefficient between the two CCDs. \\
{\it Variation of calibration coefficient during a TF scan:} given that 
the airmass variation from one observation to the next is taken into account 
in obtaining the \ktf, this variation was not expected for a photometric 
night. The rms dispersion for the 4 stars that are used to calculate \ktf\ 
is around 3\%, which is much smaller than the overall variation (9\% over the 
mean value). The trend of the variation is identical for the two scans.\\
{\it Variation of calibration coefficient between the two CCDs:} OSIRIS uses 
two CCDs, along with 2 separate electronics to cover the total field of view. 
The differences in the background levels between the two CCDs are easily 
noticeable. From our analysis, we find that there is a $6\pm3$\% difference 
between the efficiencies of the two CCDs for the H$\alpha$ scan. 
The most likely reason for this difference is the value of the gain parameter, 
which is 0.95 electrons/ADU for the right CCD and 0.91 electrons/ADU for the 
left CCD (private communication from GTC technical staff) during the 
H$\alpha$ scan.
\begin{table}
 \centering
 \caption{\label{Tab:CalCoef}Derived values of calibration coefficients}
  \begin{tabular}{lcc}
  \hline
 TF Scan & \ktf  & Units \\
\hline
H$\alpha+$[\nii]-P1 & $6.54\pm0.27$ & $10^{-18}$~erg\,s$^{-1}$\,cm$^{-2}$/(ADU/s) \\ 
H$\alpha+$[\nii]-P2 & $6.63\pm0.28$ & $10^{-18}$~erg\,s$^{-1}$\,cm$^{-2}$/(ADU/s) \\ 
\sii-P1            & $7.06\pm0.14$ & $10^{-18}$~erg\,s$^{-1}$\,cm$^{-2}$/(ADU/s) \\ 
\sii-P2           & $7.06\pm0.21$ & $10^{-18}$~erg\,s$^{-1}$\,cm$^{-2}$/(ADU/s) \\ 
\sii-P3          & $7.27\pm0.23$ & $10^{-18}$~erg\,s$^{-1}$\,cm$^{-2}$/(ADU/s) \\ 
\hline
\end{tabular}\\
\end{table}
\begin{figure*}
\begin{center}
\centerline{\epsfig{file=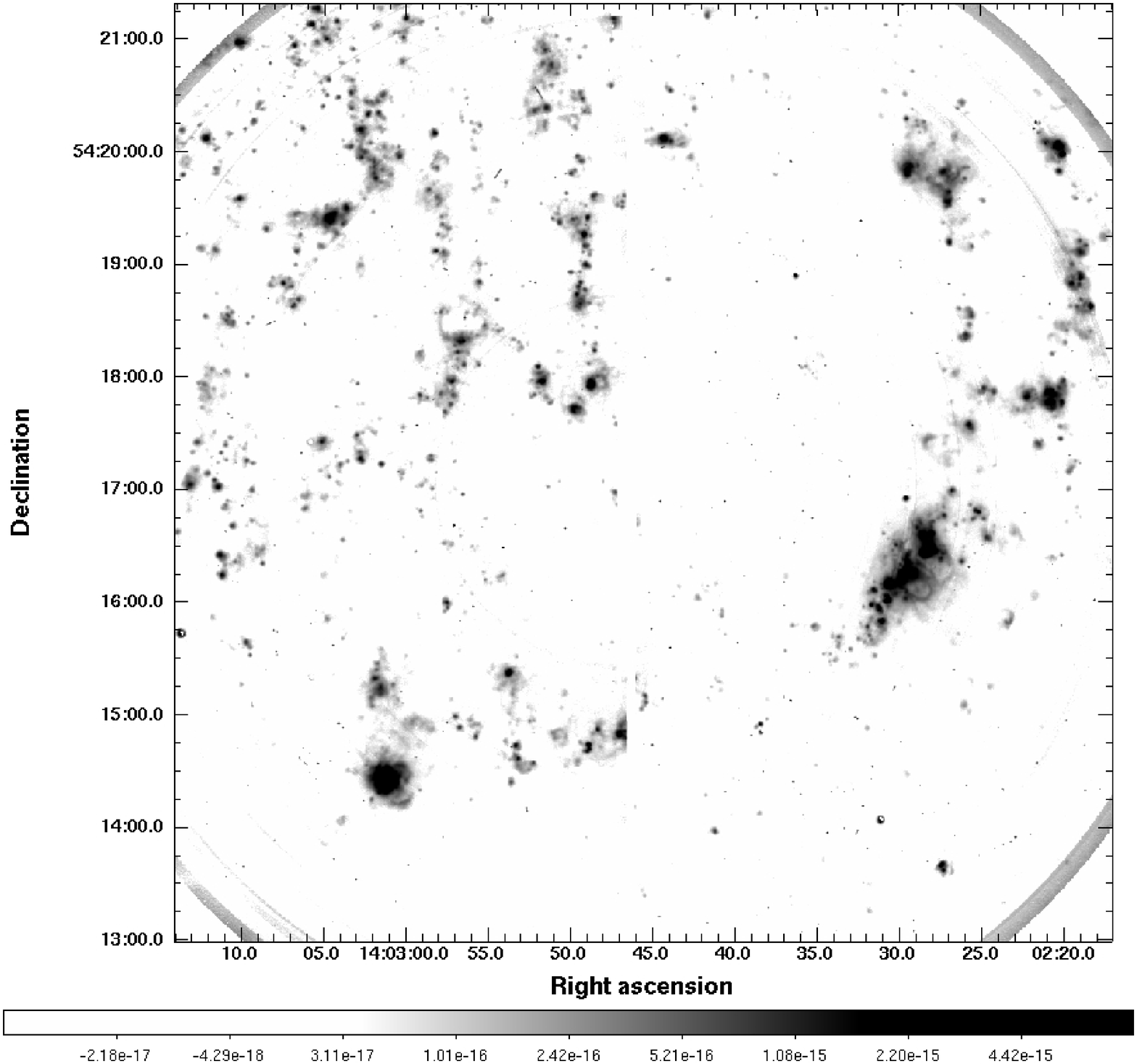, height=0.9\hsize}}
\caption{Reconstructed H$\alpha$ image of the south-west part of M101.  
Several \hii\ regions, including the massive \hii\ complex NGC~5447
($\alpha=14:02:30, \delta=54:16:15$),
as well as several filamentary structures can be seen in this image.
The brightness of the H$\alpha$ structures is shown by the gray bar
at the bottom of the image in units of erg\,s$^{-1}$\,cm$^{-2}$\,arcsec$^{-2}$, 
spanning a range from $-3\sigma$ to $512\sigma$ in logarithmic scale.
\label{fig_image}}
\end{center}
\end{figure*}

When these systematic variations in \ktf\ are taken into account, the 
calibration coefficients obtained using different stars in different CCDs 
and in different TF images, agree to within 3\% of each other. 
The mean calibration coefficients obtained from the stars on the right CCD 
of the first image of the H$\alpha$ and [\sii] scans are given in 
Table~\ref{Tab:CalCoef}.
The difference in the calibration 
coefficients for the two H$\alpha$ scans is of the order of 1\%, whereas
for the [\sii] scans this difference is around 3\%.
The mean difference in the calibration coefficients for 
the H$\alpha$ and [\sii] scans is $\sim7$\%. The \ktf\ factors in the 
reconstruction equations take into account both these
systematic variations, thus allowing us to combine the monochromatic images 
from different images of a scan.

\begin{figure}
\begin{center}
\vspace*{-2cm}
\includegraphics[scale=.47]{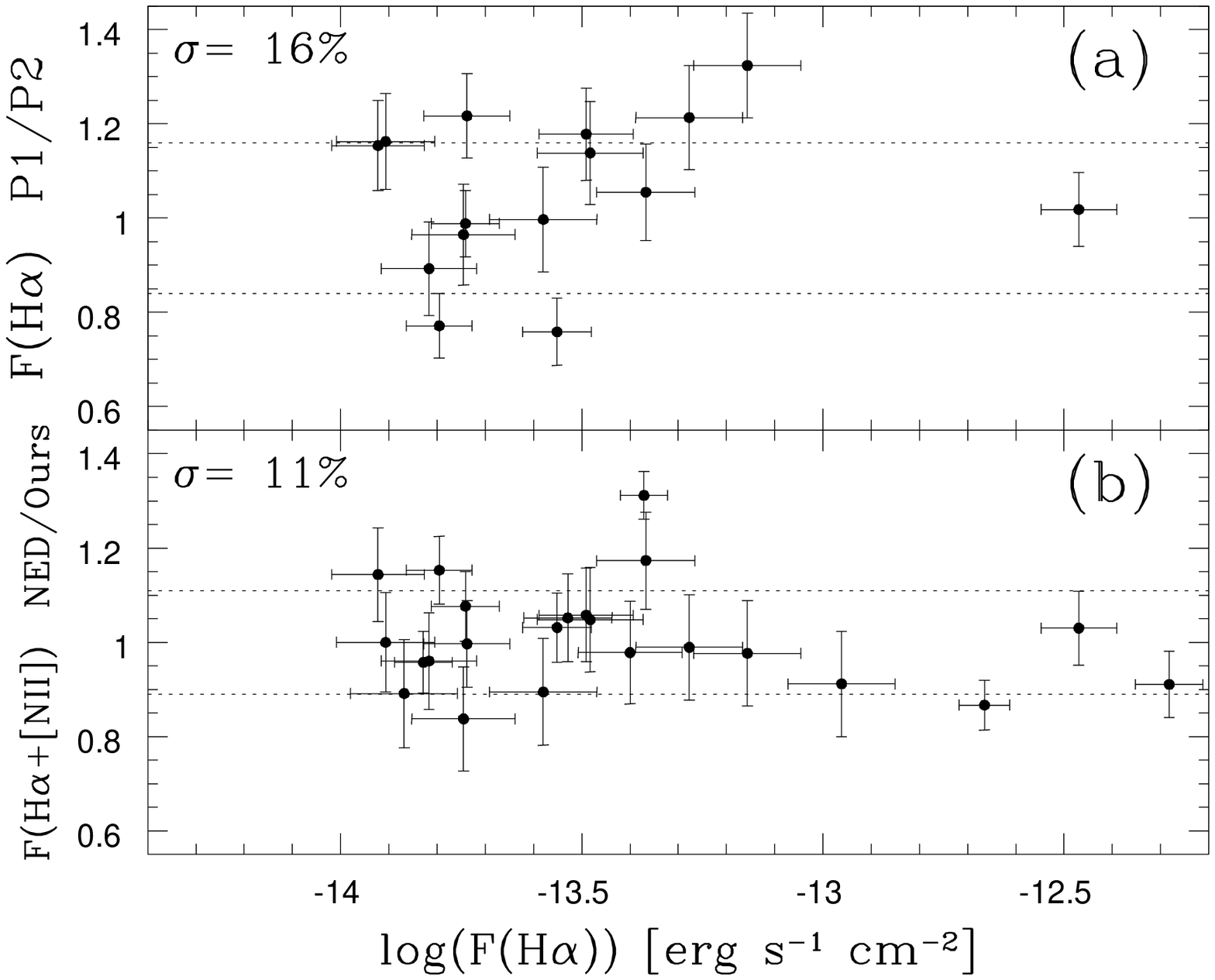}
\caption{
Relative errors on the H$\alpha$ fluxes of 23 \hii\ regions in M101.
(a) Ratio of the H$\alpha$ fluxes measured on the reconstructed 
image for the dithered position P1 to those for position P2; (b) ratio 
of the \hii\ region fluxes measured on an H$\alpha$+[\nii] image taken from NED 
to those on our images, both plotted against our H$\alpha$ fluxes. The dotted 
horizontal lines denote 1$\sigma$ scatter over the mean ratio. 
\label{fig_phot_comp1}}
\end{center}
\end{figure}

\begin{figure}
\begin{center}
\includegraphics[scale=.47]{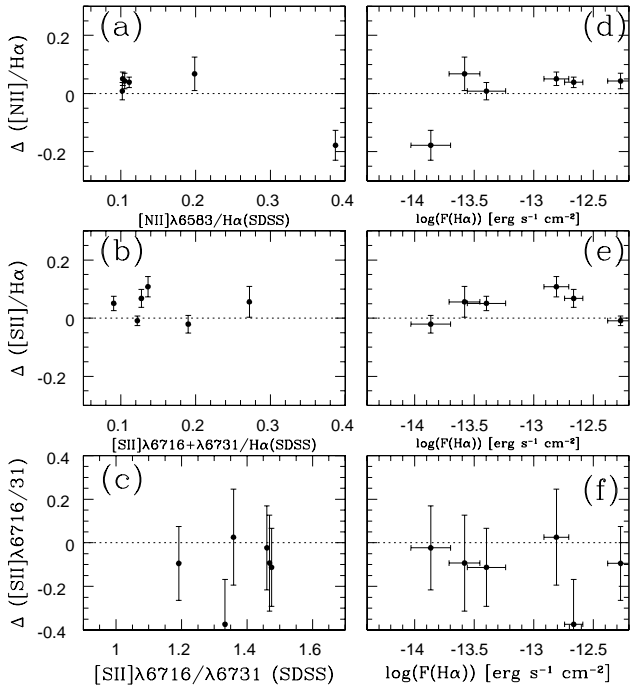}
\caption{
Comparison of important diagnostic ratios from our reconstructed images
with those obtained using SDSS spectra for 6 regions in common.
In the left panels the difference between the ratios 
(e.g. $\Delta$([\nii]$/{\rm H}\alpha$) = ([\nii]$\lambda 6583/{\rm H}\alpha{\rm )}_{\rm ours} - $ ([\nii]$\lambda 6583/H\alpha {\rm )}_{\rm sdss}$)
is plotted against the SDSS ratios, whereas in the right panels, the difference 
between the ratios is plotted against our H$\alpha$ fluxes. In general, the 
diagnostic ratios are reproduced within the ranges allowed by the 
estimated errors. See text for more details.
\label{fig_sdss_comp}}
\end{center}
\end{figure}

\begin{figure}
\begin{center}
\vspace*{-1cm}
\includegraphics[scale=.47]{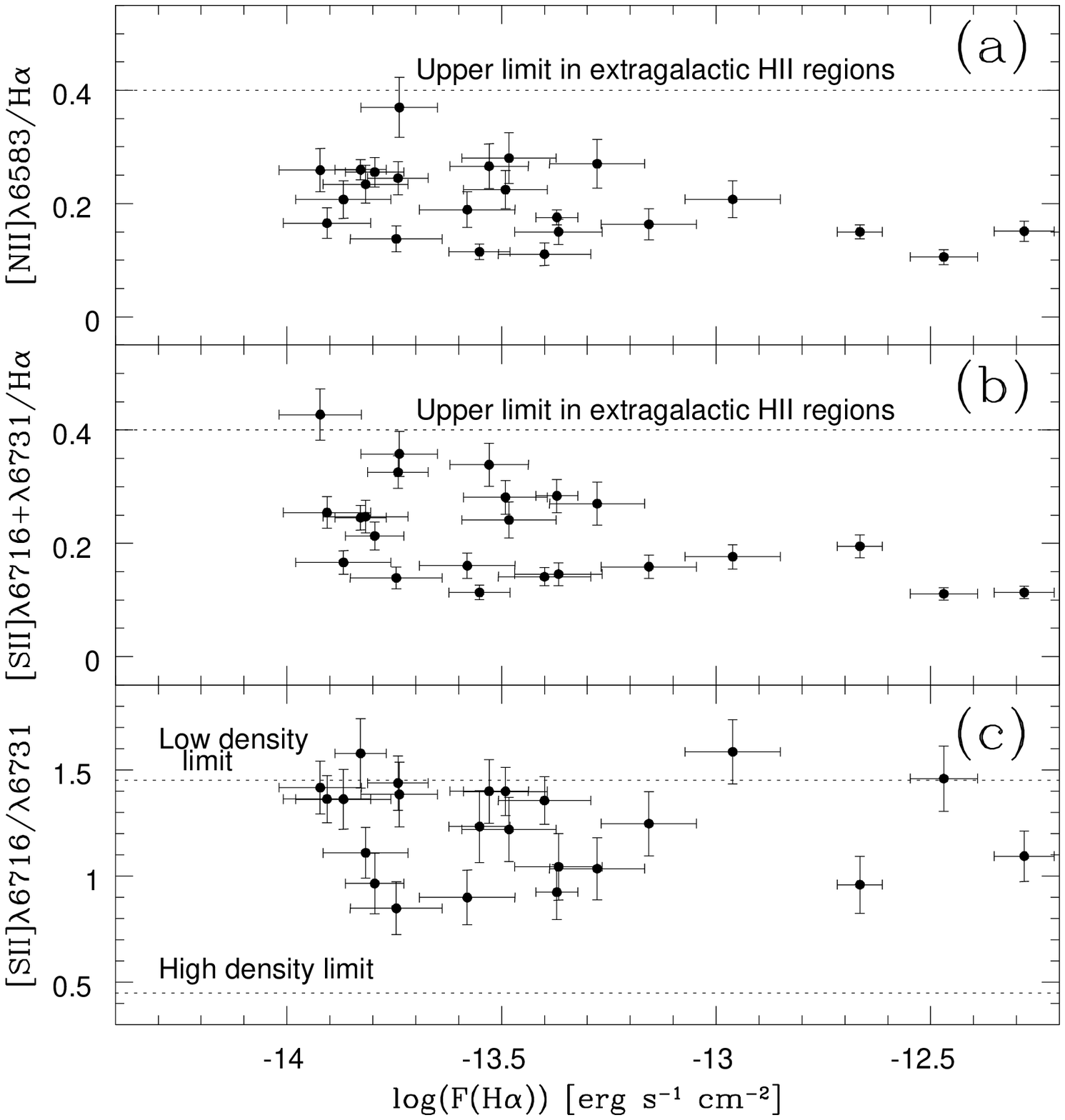}
\caption{
Flux ratios of nebular diagnostic lines plotted against the 
H$\alpha$ fluxes for the 23 \hii\ regions of Fig.~\ref{fig_phot_comp1}. 
In panels (a) and (b), we show [\nii]$\lambda$6583/H$\alpha$ 
and [\sii]$\lambda6716+\lambda6731$/H$\alpha$, whereas in panel (c) 
we show the density-sensitive ratio [\sii]$\lambda6716/\lambda6731$.
The upper limits observed in extragalactic \hii\ regions from
\citet{Deni02} are indicated by the horizontal dotted lines in the
top two panels, whereas the theoretically valid range for the [\sii]
line ratios from \citet{Oste06} is shown in panel (c). All the regions have
observed ratios in the range expected for \hii\ regions, illustrating 
the capability of TF imaging for obtaining these diagnostic ratios.
\label{fig_phot_comp2}}
\end{center}
\end{figure}

\begin{figure}
\begin{center}
\includegraphics[scale=.47]{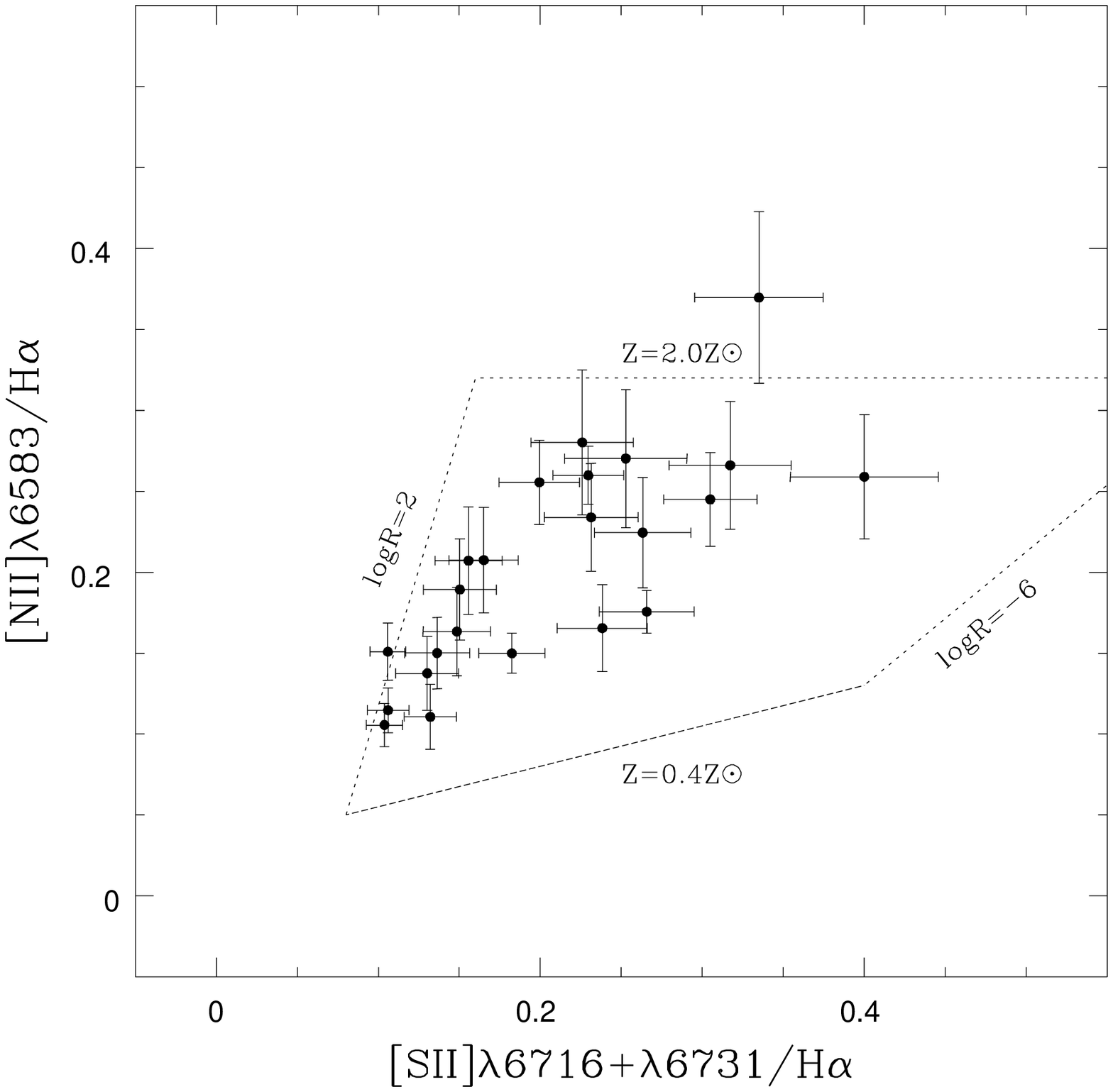}
\caption{
Nebular diagnostic diagram involving [\nii]$\lambda$6583/H$\alpha$ and
[\sii]$\lambda6716/\lambda6731$ for \hii\ regions in M101. All the observed
points lie within locii of models of \citet{Dopi06} defined by metallicities 
between $0.4Z\odot$ and $2Z\odot$, and the logarithmic ionization parameter 
$\log R$ between 2 (low pressure regions) and $-6$ (high pressure regions).
\label{fig_phot_comp3}}
\end{center}
\end{figure}

\section{Comparison of reconstructed images}

In Fig.~\ref{fig_image}, we show the reconstructed  H$\alpha$ emission-line 
image of  the south-west  part of  M101. The H$\alpha$  line being a good
tracer of ionized  gas, this image shows the  location of \hii\ complexes and
ionized filaments. It can be seen  from the image that we are able  to detect 
\hii\ regions in the  entire field of  view. 
The $5\sigma$ surface brightness limit
in this image is $8\times10^{-17}$~erg\,cm$^{-2}$\,s$^{-1}$\,arcsec$^{-2}$.
All the regions seen
in  the image  are  recovered in  the two independent  scans. In  the
following paragraphs, we compare the fluxes of recovered emission lines of \hii\  
regions in  these two independent  scans.  Our H$\alpha$ fluxes  are also
compared   with  the   H$\alpha$  fluxes   obtained   using  traditional
narrow-band imaging  technique.

We  performed photometry of
23 \hii\ regions using an aperture radius of 4\arcsec\ on our reconstructed 
H$\alpha$, [\nii]$\lambda6583$,  [\sii] images  and also on  an  H$\alpha+$[\nii]
image  that was  obtained  in the  traditional  way using  narrow-band
filters (downloaded from NED: Telescope: KPNO Schmidt; Observers: 
B. Greenawalt, and R. Walterbos). The selected regions are among the brightest
regions in the FoV. 

With a sampling of $\Delta\lambda_{\rm c}=20$~\AA\ for our dataset, $\eta_{\rm line}$ should be 
at least 0.45 in order to obtain an image without data gaps. 
However, we found systematically larger dispersion in fluxes for regions 
with response values less than 0.5. Therefore we carried out the reconstruction 
with $\eta_{\rm line}=0.5$. As a result, eight regions
have data in only one of the dithered scans.
In Fig.~\ref{fig_phot_comp1}a, we show a comparison of the fluxes of the
15 \hii\ regions 
whose fluxes could be measured on the reconstructed images from both 
the scans.
The mean ratio of the H$\alpha$ fluxes is unity over 2 orders of 
magnitude in flux, confirming the accuracy of the method we adopted for
the flux calibration. The rms dispersion of the H$\alpha$ fluxes of the same 
region obtained on images for the two scans is 16\%. This value is within 
the flux errors expected for the 1--2~\AA\ error in $\lambda_{\rm c}$.

In order to check whether the reconstruction process rightly reproduces 
relative fluxes of different regions over the entire FoV, 
we plot the ratio of the H$\alpha+$[\nii]$\lambda6583$ fluxes measured from 
our reconstructed images (P1 and P2), and 
that from traditional narrow-band filters in Fig.~\ref{fig_phot_comp1}b.
Our fluxes are given as the average  
when data are available from both scans. As the intention here is 
a comparison of relative fluxes, we set the mean value of the flux
ratio to unity. There is $\sim11$\% scatter on this mean value, which is
marginally better than that between P1 and P2, as expected due to the use
of averaged fluxes. There is a marginal trend for the
mean ratio to be $\sim5$\% different between the bright and faint regions,
which is nevertheless smaller than the scatter. 
We did not find any trend of these two ratios against the distance of the
region from the optical center.

Three sources of error are included in calculating the sizes of
the error bars plotted in Fig.~\ref{fig_phot_comp1}. They are 
(1) photon noise of the object, (2) the error in the subtraction of the sky value and
(3) the error in the recovered flux due to an error in $\lambda_c$. 
The last of these errors, which is discussed in detail in \S5,
dominates for the 23 regions for which we performed photometry, contributing 
around 10\% for the majority of the regions.

As a second test of the reliability of the flux ratios obtained using the
TF images, we compare the flux ratios of emission
lines relevant for diagnostic diagrams with those values obtained using spectra from the literature for the same regions.
Seventeen \hii\ regions in M101 were observed spectroscopically by SDSS 
\citep{Shol07}. Six of these lie within the usable FoV of our images (radial 
distance from the optical center $\lesssim3.75^\prime$). 
Flux ratio of nebular diagnostic lines from our reconstructed images are 
compared with those obtained from the SDSS spectra in 
Fig.~\ref{fig_sdss_comp}. 
Our fluxes were obtained over apertures of $4\arcsec$ radius at the coordinates
associated with the SDSS spectra.
The adopted apertures, though almost 3 times bigger than the fiber sizes of 
SDSS spectra (3\arcsec\ diameter), are the minimum area over which reliable fluxes 
can be measured in our images. Given that the spectroscopic ratios of giant 
\hii\  regions are not expected to vary much with aperture size, our relatively 
bigger apertures are not expected to introduce additional errors.
The difference in the flux ratios between ours and SDSS values are plotted 
both against flux ratios (in the left panel) and H$\alpha$ fluxes
(in the right panel). The plotted error bars take into account all the
errors discussed in the paragraph above. The errors in the spectroscopic 
ratios are expected to be almost negligible, and hence we did not include
these errors in our analysis. The majority of the points lie close to the
horizontal line within the plotted errors, indicating that the ratios of
lines relevant for diagnostic purposes can be obtained using TF imaging. 
There are a few ratios that deviate from the spectroscopic ratios by more
than the estimated errors. The most important of them is the right-most
point in panel (a). Apart from being the faintest in H$\alpha$, 
the spectroscopic [\nii]$\lambda6583$/H$\alpha$ value for this region is 0.4,
a value that is too high as compared to the assumed value of 0.1 in 
Eqn~\ref{eqn_recon8}. 
This is the most likely reason for the large deviation of this region.
Kinematics of individual regions with radial velocities 
$>25$~km\,s$^{-1}$ (i.e. 0.5~\AA\ around the H$\alpha$ wavelength) can 
also be responsible for the observed deviations of some of 
the points. The errors in the measured radial velocities of the regions 
from SDSS spectra do not permit us to carry out a more elaborate analysis
of the deviations of individual regions.

In summary, the ratios recovered from the TF images for individual regions
have larger errors than the corresponding spectroscopic ratios. 
Nevertheless, the capability of TF images to derive such ratios for 
large number of regions, makes TF imaging a scientifically attractive 
option, as we illustrate below using the flux ratio of 23 \hii\ regions in M101.

In Fig.~\ref{fig_phot_comp2}, we plot the
[\nii]$\lambda$6583/H$\alpha$, [\sii]$\lambda6716+\lambda6731$/H$\alpha$, 
and the [\sii]$\lambda6716/\lambda6731$ ratio of \hii\ regions against their 
H$\alpha$ fluxes in panels (a), (b) and (c), respectively. In each of these 
panels, we indicate the observed/theoretical ranges of these ratios in \hii\ 
regions, which cover $\sim10$ times the errors on the ratios. This relatively
large dynamic range, combined with the intrinsic multi-object capability of 
TF imaging, makes it competitive with traditional spectroscopic observations.

The [\nii] and [\sii] lines originate in low ionization zones in \hii\ regions. 
Their ratio to the H$\alpha$ flux depends not only on the ionization 
parameter, but also on the abundance of these ions. The utility of these 
ratios for diagnostic purposes has been discussed in \citet{Bald81}.
In Fig.~\ref{fig_phot_comp3}, we show the values for these ratios expected 
for a range of ionization parameters ($R$) and metallicities ($Z$), using
the models of \citet{Dopi06}. The observed values of these ratios in our
selected 23 regions lie within the range of model values. The regions having
high values of [\nii]/H$\alpha$ have most likely high
nitrogen abundances, as was found by \citet{Deni02} for \hii\ regions in a
sample of nearby galaxies. Thus, OSIRIS TF imaging is very promising for
the study of nebular line ratio diagnostics of nearby large galaxies.

\section{Summary}

In this work, we have explored the capability of tunable filter imaging
with the OSIRIS instrument at the 10.4-m GTC, using real data.
The changing wavelength across the field and the non-flat
functional form of the response curves makes it essential to have a 
sophisticated analysis package to completely take advantage of the 
relatively large field of view of the instrument. With the set-up that we
have used to observe M101, we were able to obtain monochromatic images
in the emission lines of H$\alpha$, [\nii]$\lambda6583$ and the [\sii] doublet.
We demonstrate that line fluxes and their ratios for 
prominent \hii\ regions can be obtained to better than $\sim$15\%, which is 
basically limited by the current 1~\AA\ uncertainty in setting the central 
wavelength of the TF. 
Though these errors are much larger than the spectroscopic ones, 
the multi-object capability of TF imaging, combined with the relatively 
high sensitivity of the flux ratios to variations in density, excitation 
and metallicity, makes TF imaging an attractive option for investigating 
the point-to-point variation  of these physical quantities in galaxies.
We also demonstrate that the emission-line maps
can be flux-calibrated to better than 3\% accuracy, using the $griz$ SDSS
magnitudes of in-frame stars, and the spectral database of SDSS, 
without the need to invest extra telescope time to perform the
photometry of spectrophotometric stars. 

\acknowledgments

It is a pleasure to thank an anonymous referee, whose thoughtful comments
helped us to improve the original manuscript.
We would also like to thank the GTC/OSIRIS staff members, especially Antonio 
Cabrera, for the support provided for this project.
This work is partly supported by CONACyT (Mexico) research grants
CB-2010-01-155142-G3 (PI:YDM), CB-2011-01-167281-F3 (PI: DRG), 
CB-2005-01-49847 and CB-2010-01-155046 (PI: RJT), and  CB-2008-103365 (PI: ET).
This work has been also partly funded by the Spanish MICINN, Estallidos 
(AYA2010-21887, AYA2007-67965) and Consolider-Ingenio (CSD00070-2006) grants.

\end{document}